\documentclass{svjour2}                    
\smartqed  
\usepackage{graphicx}
\begin{document}

\title{Convergence of the enhancement of the effective mass under pressure and magnetic field in heavy-fermion compounds: CeRu$_2$Si$_2$, CeRh$_2$Si$_2$, and CeIn$_3$
}


\author{J. Flouquet \and D. Aoki \and W. Knafo \and G. Knebel \and T.D. Matsuda \and S. Raymond \and C. Proust \and C. Paulsen \and P. Haen
}


\institute{J. Flouquet \and D. Aoki \and G. Knebel \and T.D. Matsuda \and S. Raymond \at
              SPSMS, UMR-E 9001, CEA-INAC/ UJF-Grenoble 1, 17 rue des Martyrs, 38054 Grenoble Cedex 9, France
 \\
              Tel.: +33-43878-5423 \\
              Fax: +33-43878-5096\\
              \email{jacques.flouquet@cea.fr}           
           \and
           W. Knafo \and C. Proust \at
             Laboratoire National des Champs Magn\'{e}tiques Intenses, UPR 3228, CNRS-UJF-UPS-INSA, 143 Avenue de Rangueil,
				31400 Toulouse, France\\
            \and
           C. Paulsen \and P. Haen
			\at Institut N\'{e}el, CNRS / UJF Grenoble, BP166, 38042 Grenoble Cedex 9, France
}

\date{Received: date / Accepted: date}

\maketitle

\begin{abstract}
Emphasis is given on the observation of a convergence to a critical value of the effective mass of a heavy fermion compound by tuning it through a quantum instability either by applying pressure or magnetic field from an antiferromagnetic (AF) to a paramagnetic (PM) ground state. Macroscopic and microscopic results are discussed and the main message is to rush to the discovery of an ideal material whose Fermi surface could be fully observed on both sides of each quantum phase transition.

\end{abstract}

Due to the weakness of the renormalized parameters, such as the effective Fermi temperature $T_F$ in heavy fermion compounds (HFC), it is possible to tune them with moderate values of pressure ($p$) or magnetic field ($H$)   from a long range antiferromagnetic (AF) ground state to a paramagnetic (PM) one at a critical pressure $p=p_c$ or a critical field $H=H_c$ \cite{Flouquet2005}. Under pressure the main phenomena can be considered to be governed by the collapse of the AF order parameter. At low temperature and under magnetic field, often the achievement of a significant high magnetic polarization near $H_c$ ends up in a polarized paramagnetic (PPM) phase with a marked crossover on warming from the low field paramagnetic phase. Often, for fields close to the critical field $H_c$, the magnetic polarization of the 4$f$ centers reaches typically 20\% of the full moment. Thus, ferromagnetic interactions must certainly play a major role. Assuming that the 4$f$ electrons are itinerant, the difference between the majority and minority spin bands should have consequences on the Fermi surface.

\begin{figure}
\includegraphics[width=0.75\textwidth]{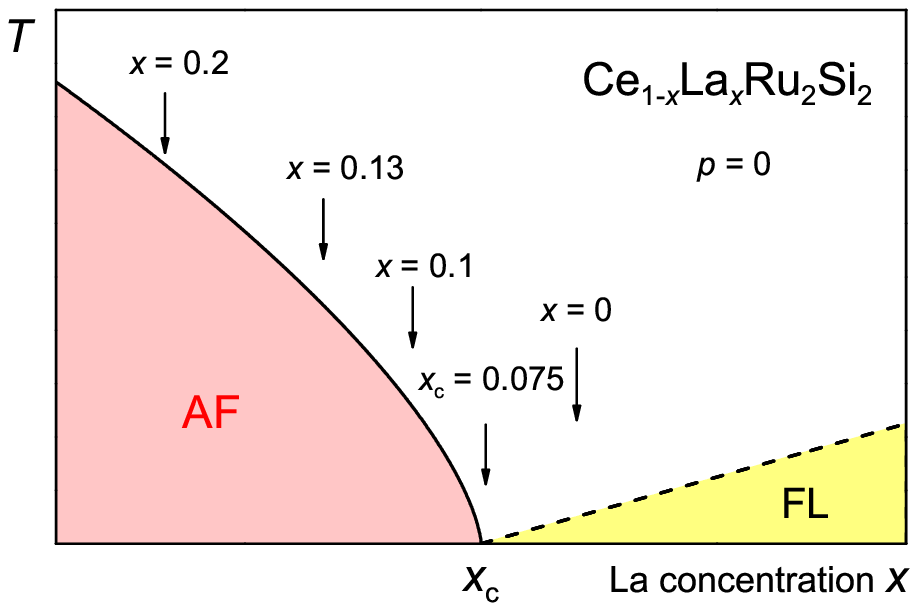}
\caption{($T, x$) phase diagram of Ce$_{1-x}$La$_x$Ru$_2$Si$_2$ which is representative of the ($T, p$) phase diagram of an AF HFC. The AF order is suppressed at $p_c$ ($x_c$ in case of doping). The location of CeRu$_2$Si$_2$ and Ce$_{1-x}$La$_x$Ru$_2$Si$_2$ for different La concentrations at $p=0$ are indicated. }
\label{fig:1}
\end{figure}

On the paramagnetic side of the phase diagram (see Fig. \ref{fig:1}) the vicinity of the critical pressure $p_c$ is characterized by a large electronic Gr\"uneisen parameter $\Omega (T \to 0)$ \cite{Benoit1980,Takke1981}. The Gr\"uneisen parameter $\Omega (T)$ is defined by the ratio of the  thermal expansion $\alpha$ and the specific heat $C$ multiplied with the ratio of the  molar volume $V$ and the compressibility $\kappa$ ($\Omega (T) = \frac{\alpha}{C} \times \frac{V}{\kappa}$). It has been observed that a constant value of $\Omega (T)$  is only approached at very low temperatures \cite{Benoit1980,Takke1981}. Basically, the continuous increase of $\Omega (T)$ on cooling is a direct macroscopic evidence of a large non Fermi liquid domain in temperature. In this regime, the free energy $F$ is not controlled by a single energy scale $T^\star$ and  $F$ cannot be reduced to a simple expression $F = T \Phi (T/T^\star )$ which would imply $\Omega (T) = \Omega (0) = - \partial \log T^\star / \partial \log V$.

In many HFC such as CeNi$_2$Ge$_2$ \cite{Kuechler2003}, $\Omega (T \to 0)$ seems to diverge at $p_c$ . Figure 2 shows the temperature dependence of $\Omega$ of CeRu$_2$Si$_2$ at $p=0$ \cite{Flouquet2005,Lacerda1989}. The specific interest of CeRu$_2$Si$_2$ is that at $p=0$ it is located slightly above the ''effective'' negative critical pressure ($-p_c = 0.2 - 0.5$ GPa) \cite{Payer1993}. The Ising character of the uniform magnetization leads to clear first order metamagnetic phenomena, when by expanding the volume, long range AF order is recovered. Such lattice dilatation is realized by substitutions on the Ce, or Si sites either by La \cite{Fisher1991}, or Ge \cite{Haen1999}, respectively, whereas the substitution of Ru by Rh is not iso-electronic but induces AF  ordering too \cite{Sekine1992}. 
Here, we will first focus on the Ce$_{1-x}$La$_x$Ru$_2$Si$_2$ series. At ambient pressure the critical concentration at which the N\'eel temperature $T_N$ vanishes is close to $x_c = 0.075$ \cite{Fisher1991,Knafo2009}. On the AF side, at a concentration $x=0.2$ ($T_N = 6$ K), sweeping the magnetic field at zero pressure leads  to two successive metamagnetic transitions at $H_a$ and $H_c$, corresponding to phase transitions between two AF structures (at $H_a$) and between AF and PM phases (at $H_c$) \cite{Fisher1991}. Furthermore, for this La concentration $x = 0.2$, antiferromagnetism collapses at a pressure $p_c = 0.4$ GPa  while the metamagnetic transition terminates at a critical end point $H^\star_c=3.5$ T \cite{Flouquet2005,Haen1996}. Entering in the PM side, the proximity of the critical end point at $H^\star_c$ is felt by the occurence of a sharp pseudo-metamagnetic transition at $H_M$ (see Fig. \ref{fig:3}).
The aim of the present article is to focus on the interplay between the pressure and field instabilities with a special emphasis on the mass enhancement at the critical pressure, $m^\star (p_c)$, or at the critical field,  $m^\star (H_c)$ or $m^\star (H_M)$. Our attention that the mass enhancement may be the same at the critical points $p_c$, $H_c$ or $H_M$ emerged from recent measurements made on CeRh$_2$Si$_2$ which is a HFC situated deep inside the AF phase at $p=0$ ($T_N = 36$ K, $H_c = 26$ T) \cite{Knafo2010}, but where AF order is rapidly suppressed under pressure. For CeRh$_2$Si$_2$ a PM  ground state is realized already at $p_c \approx 1$ GPa \cite{Araki2002,Villaume2007}. Furthermore, the merging of $m^\star (H_c)$ and $m^\star (p_c)$ can be pointed out from former studies of  heavy fermion systems, as that of CeIn$_3$ either by pressure \cite{Knebel2001} or field sweep \cite{Silhanek2006}. Due to the weak magnetic anisotropy of the CeIn$_3$ cubic lattice by contrast to the large anisotropy of the CeRu$_2$Si$_2$ tetragonal lattice, $H_c$ collapses at $p_c$ in CeIn$_3$ \cite{Purcell2009}.

Thus the intercept of the crossover line $H_M$ with the critical AF line $H_c$ depends on specific conditions like the Fermi surface topology, the Ising or Heisenberg character of the local magnetization, the anisotropy of the intersite interactions, the AF wavevector, or the interplay between the AF and the Kondo fluctuations. For CeRu$_2$Si$_2$, three different wavevectors $k_1 = (0.31, 0, 0)$, $k_2 = (0.31, 0.31, 0)$ and $k_3 = (0, 0, 0.35)$ are hot spots \cite{Kadowaki2004}. For La doping antiferromagnetism develops in a transverse mode at the ordering vector $k_1$ while Rh doping series order at $k_3$ in a longitudinal mode. For both series the sublattice magnetization is aligned along the $c$ axis. This difference in the magnetic ordering is associated in the La series by the glue of $H_M$ to $H_c$ while in the Rh series a complete decoupling between $H_M$ and $H_c$ has been observed \cite{Sekine1992}. In the Rh-doped series, $H_c$ seems to collapse with $T_N$. When $H_M$ overpasses $H_c$ whatever is $p$, $H_M$ is mainly associated with the strength of the Kondo magnetic field $H_K=k_BT_K/g\mu_B$ required to quench significantly the Kondo effect. Below, a focus is made on the situation where $H_M$ touches $H_c$ below $p_c$ at finite temperature. In this case, $H_M$ is the combined result of short range intersite correlations and local fluctuations. Basically, it is the extension above $p_c$ of the magnetic critical end point $H^*_c$.

From the temperature dependence of the specific heat (Fig. \ref{fig:4}) and also from inelastic neutron scattering response \cite{Kambe1996,Knafo2009,Raymond2010}, CeRu$_2$Si$_2$ appears to be well described by the spin-fluctuation theory of Hertz, Millis, and Moriya  (HMM) \cite{Hertz1976,Moriya1995,Millis1993}, which was first developed and tested to describe of weak antiferro- or ferromagnetism in 3$d$ intermetallics \cite{Lonzarich1989}. For HFCs a novelty is the necessity to use an already renormalized Fermi temperature $T_F$ which is crudely associated with the Kondo temperature of the 4$f$ ions \cite{Moriya1995}. This scheme seems now well established by the recent collection of inelastic neutron scattering data on the Ce$_{1-x}$La$_x$Ru$_2$Si$_2$ series for different La concentrations ranging from $x = 0.2$ to $x=0$, i.e. covering the sweep from the AF to the PM ground states \cite{Knafo2009}. As shown in Fig. \ref{fig:5}, the real part of the susceptibility $\chi '(Q_1,T)$, measured at the momentum transfer $Q_1$ characteristic of the AF order parameter (at the wavevector $k_1$), has a maximum value at the critical temperature $T_N$, which collapses at the critical concentration $x_c$.  Oppositely, no maximum can be detected at a momentum transfer $Q_0$ far from the AF hot spots. Basically, the temperature range where $\chi (Q_0)$ saturates as well as the amplitude of $\chi (Q_0)$ is governed by the volume dependence of $T_K$.  This study has shown that fluctuations of the AF order parameter govern the transition from paramagnetism to antiferromagnetism in Ce$_{1-x}$La$_x$Ru$_2$Si$_2$.

A key hypothesis in the spin-fluctuation approach is the invariance of the Fermi surface through $p_c$. Fortunately due to the high quality of the crystals, CeRu$_2$Si$_2$ is one of the rare cases of HFC where the Fermi surface has been fully determined \cite{Aoki1992,Aoki1995,Takashita1996,Matsumoto2008} and which allows a serious comparison with band structure calculations. The topology of the Fermi surface appears well described in a model where the 4$f$ electrons are treated as itinerant and the crystal field is taken into account \cite{Suzuki2010}. Due to the improvements in the accuracy of angular resolved photoemission spectroscopy (ARPES) it was recently demonstrated that this PM topology of the Fermi surface is mainly preserved in the AF situation of CeRu$_2$(Si$_{0.82}$Ge$_{0.18}$)$_2$ (see Fig. \ref{fig:6}) in sharp contrast with a 4$f$ localized treatment where the Fermi surface should look like that of LaRu$_2$Si$_2$ \cite{Okane2009}. Experiment and theory \cite{Hoshino2009,Miyake2006} support an itinerant picture of the 4$f$ electron whatever is the ground state (AF, PM or PPM). Of course, at the magnetic ordering, a Fermi surface reconstruction might be governed by the new AF Brillouin zone.

To summarize these studies with chemical doping, which can be considered as equivalent to studies under pressure, macroscopic and also microscopic measurements seem  to support strongly a spin-fluctuation approach. However, from the temperature variation of the specific heat divided by the temperature shown in Fig. \ref{fig:4}, a conventional behavior of the AF fluctuations with a singularity of the extrapolated value $\gamma =(C/T)_{T \to 0 K}$ in $\sqrt{p-p_c}$ or $\sqrt{x-x_c}$ does not seem to be reproduced. For example, at a La concentration $x = 0.13$ (which corresponds to a pressure of $p_c - 0.2$ GPa), $C/T$ shows a sharp jump at $T_N \approx 4$ K, but on cooling there is a large temperature window where $C/T$ remains almost constant near a critical value $\gamma_c \approx 600$ mJ mole$^{-1}$ K$^{-2}$, which is exactly the low temperature value of $C/T$ at the critical concentration $x = x_c = 0.075$ \cite{Fisher1991,Raymond2010}. For $x= 0.13$, the transition at $T_L \approx 0.7$ K corresponds to a  change in the magnetic structure which opens out in a decrease of $C/T$ at lower temperatures. This phenomena is very clear at $x = 0.2$. Surprisingly, focus on AF quantum criticality has been made mainly on the PM side ($p=p_c + \epsilon$) and attempts to describe the AF side have been completely omitted. Another negative point for the HMM approach is that the usual link between the dependence of $T_N$ with the size of the sublattice magnetization ($T_N \sim M^{3/2}$) has not  been verified \cite{Flouquet2005,Raymond2007}. Even on the PM side, a tiny ordered moment $M_0 \approx 0.01 \mu_B$ survives in CeRu$_2$Si$_2$ \cite{Amato1993}. This residual antiferromagnetism is up to now believed to originate from lattice imperfections which may create locally pressure gradients of a few kbar and thus would play the role of nucleation centers for the occurrence of residual AF droplets. Finally, to our knowledge, no divergence of the magnetic correlation length at $p_c$ or $x_c$ has been reported from the experiment for any heavy fermion compound \cite{Flouquet2005}. So the definitive proof of a second order quantum criticality is missing.
A clear signature is that $\gamma$ reaches a critical value at $p_c$ corresponding to a critical value of the average effective mass $m^\star_c = m^\star(p_c)$.

The interplay of the different mechanisms involved in the field restoration of the PM phase from an AF ground state is well shown in Fig. \ref{fig:7}a where the Ce$_{1-x}$La$_x$Ru$_2$Si$_2$ antiferromagnet of concentration $x=0.1$ is considered. Here the differential susceptibility $\partial M/\partial H$ of the magnetization is shown as a function of $H$ at different temperatures \cite{Fisher1991}. On cooling below $T_N \sim 3$ K, the two first order metamagnetic transitions at $H_a$ and $H_c$ emerges clearly. However, just above $H_c$ a shallow maximum persists at $H_M$. On warming the differentiation between $H_c$ and $H_M$ increases. Above $T_N$ only the broad maximum at $H_M$ persists. $H_c$ is characteristic of the spin flip of the static magnetization \cite{Mignot1991} while $H_M$ is governed by the interplay of FM, AF, and local spin dynamics \cite{Flouquet2004,Sato2004}. For the PM ground state, only a pseudo-metamagnetic transition survives; furthermore its characteristic magnetic field $H_M$ corresponds to a critical value $M_c= 0.5 \mu_B$/Ce-ion of the magnetization $M$, i.e. of the magnetic polarization. It is remarkable that under pressure, even at 1 GPa above $p_c$, the pseudo-metamagnetic crossover corresponds to $M(H=H_M) = M_c$ \cite{Mignot1989}. As for $H \sim H_M$, the strength of the inelastic electronic scattering is mainly pressure invariant, so that $m^\star (H_c)$ for $p<p_c$ might be nearly equal to $m^\star (H_M)$ for $p>p_c$.

The pseudo-metamagnetic transition in CeRu$_2$Si$_2$ has been highly studied by magnetization \cite{Flouquet2005,Haen1987,Paulsen1990,Sakakibara1995}, specific heat \cite{Fisher1991,vanderMeulen1991,Kim1990,Aoki1998}, transport \cite{Kambe1996,Daou2006}, ultrasound \cite{Kouroudis1987}, NMR \cite{Ishida1998}, elastic and inelastic neutron measurements \cite{Flouquet2005,Regnault1988,Raymond1999}, as well as quantum oscillation methods \cite{Aoki1992,Aoki1995,Takashita1996}. It is worthwhile to remark that at $H=0$, AF correlations have been detected by neutron scattering up to $T_{corr}\approx 60$ K \cite{Regnault1988} pointing out that their onset occurs far above the Kondo temperature $T_K \sim 20$ K, which is assigned by simple considerations on the specific heat maximum. Under magnetic field, a sensitive tool to detect the field of the metamagnetic transition $H_M (T)$ has been the determination of the maximum of the magnetoresistance at constant temperature \cite{Haen1987}. It has been shown that $H_M$ collapses also around $T = 60$ K and that it reaches a constant value only below 1 K. This is shown in Fig. \ref{fig:8}. Thermal expansion measurements at different magnetic fields were a quite powerful tool to draw the crossover boundaries in Fig. \ref{fig:9} \cite{Paulsen1990}: below $H_M$, the ($H,T$) boundary looks like that of an AF state, but it consists in a paramagnetic phase with strong antiferromagnetic fluctuations, and above $H_M$ it looks like that of a ferromagnet, but is then a crossover to a PPM state.

The demonstration that FM fluctuations play a major role in the sharpness of the pseudo-metamagnetic phenomena has been obtained via two inelastic neutron experiments \cite{Flouquet2004,Sato2004} with the observation that close to structural Bragg reflections at $Q = (0.9, 1, 0)$ a strong field-induced softening of the FM fluctuations occurs on approaching $H_M$ (see Fig. \ref{fig:10}). In contrast,  the vanishing of the AF fluctuations under magnetic field does not occur via an AF instability but via an increase of the damping of the AF fluctuations with field. Macroscopically, the consequence of the switch under field from dominant AF to FM interactions is the increase of the Sommerfeld coefficient $\gamma = C/T$ for $T\to 0$ with magnetic field, as demonstrated in Fig. \ref{fig:11} where the singularity  at $H_M$ is quite analogous to that predicted by the AF spin-fluctuation theory. However, as we stressed above the real mechanism which drives the transition is a transfer from AF to FM fluctuations. From these studies, a key message is that at $H_M$, $\gamma (H_M)$ has almost the same value than that reached at $p_c$ so that $m^\star (p_c) = m^\star (H_M)$.

Such a convergence of the effective mass under pressure and magnetic field was recently reported for the antiferromagnet CeRh$_2$Si$_2$ at $p=0$ (Fig. \ref{fig:12}) \cite{Knafo2010}. Comparing the field and pressure variation of the $A$ coefficient of the $T^2$ Fermi-liquid term of the resistivity (which is assumed to be proportional to $\gamma^2$), led to the conclusion that $A(p_c) \sim A(H_c)$ (see Fig. \ref{fig:13}), i.e. $m^\star (p_c) \sim m^\star (H_c)$ \cite{Knafo2010}. Furthermore we stress that a marked field enhancement of $A$ starts even far below $H_c$, at a  field near $H^\star \sim 15$ T where at $T= 0$ the crossover between the PM and PPM phases would occur in absence of antiferromagnetism. Thus, an important observation is that despite the strong first order nature of the metamagnetic transition of CeRh$_2$Si$_2$ at $H_c$ (where the magnetization jumps by $\Delta M \sim 1.4 \mu_B$/Ce-ion), the field enhancement occurs far below $H_c$, i.e. at $(H^\star-H_c)/H_c \sim 0.5$. In Fig. \ref{fig:14} we have drawn schematically the $(H,p,T)$ phase diagram of CeRh$_2$Si$_2$ which is also general for other HFC. The dashed area indicates the region in the $(p, H)$ plane where FM fluctuations might play a significant role. Far above $p_v$, where valence fluctuations are expected to be large, the FM fluctuations should certainly drop. However, as will be discussed later, the FM fluctuations may be enhanced near $p_v$, even at $H \to 0$.

A further experimental evidence of the convergence of $A(p_c)$ and $A(H_c)$ can be found in the study realized on CeIn$_3$ under pressure and magnetic field and on CeIn$_{2.75}$Sn$_{0.25}$, where Sn-doping permits to lower $H_c$ down to 45 T instead of 60 T for the pure compound \cite{Silhanek2006,Purcell2009}. If we consider that a broadening is induced by doping, the data are consistent with  $m^\star (p_c) \approx m^\star (H_c)$ (Fig. \ref{fig:15}). Another similarity between the three discussed examples of CeRu$_2$Si$_2$,  CeRh$_2$Si$_2$ and CeIn$_3$ is that a large magnetic polarization is required to destroy AF correlations at the profit of FM correlations. One may hope to detect fully the Fermi surface in the AF and PPM phases by quantum oscillations techniques and even later to zoom on the Fermi surface evolution through the sharp crossover regime at $H_M$. However, very often in the experiments large parts of the Fermi surface have not been observed as for example minority spin carriers may get a too large effective mass to be detected \cite{Flouquet2005}. For example in CeRu$_2$Si$_2$, the field enhancement of $m^\star$ above $H_M$ is only observed for a few orbits and its correspondence with the average value measured via the $\gamma$ term cannot be verified \cite{Aoki1995,Takashita1996}.

Emerging from macroscopic measurements on the quite different systems CeRu$_2$Si$_2$, CeRh$_2$Si$_2$, and CeIn$_3$, a golden rule has to  be obeyed in order to find the relation $m^\star (p_c) = m^\star (H_M {\rm or}$ $H_c)$. Furthermore this equality does not require to be at the AF quantum singularity at $p_c$. For conventional magnetic materials, it is well known that applying a magnetic field generally changes the universality class of the phase transition. Thus the similarity between pressure and magnetic field tunings is not obvious. As pointed out, close to $p_c$, the AF phase has been weakly considered and discussed. A sound idea is to look more carefully to the microscopic phenomena.
The independence of the product $\Gamma_q \chi_q$ of the magnetic relaxation rate $\Gamma_q$ and the susceptibility at the wavevectors $q$ derived in the framework of a quasi-localized model \cite{Kuramoto1987} seems also obeyed in HFCs, whatever is the ground state \cite{Knafo2009}. The magnetic field induced transfer from AF to FM fluctuations may be dominated by such a rule taking into account that the first order nature of the FM instability, as well as the damping of FM fluctuations under field, prevent any collapse of $\Gamma_q$ at $q=0$ for $H = H_M$. Thus, there is a concomitant mechanism to avoid any divergence of $m^\star$ at $p_c$ and at $H_c$. Finding a $H$- or $p$-induced singularity in the density of states to explain the common convergence of $m^\star (p_c)$ with $m^\star (H_M)$ is a key issue.

Of course, an appealing route is to look deeper than before to the Fermi surface instability with the idea that the singularity has to be marked in the $p$ and $H$ evolution of the Fermi surface. Such instabilities may not require a drastic change but a topological change as discussed long time ago for the 2.5 Lifshitz transition \cite{Lifshits1960}. In CeIn$_3$ this approach was already made to explain the field-induced evolution of the Fermi surface \cite{Gorkov2006}. It was demonstrated that, via magnetic polarization, the magnetic field can lead to a logarithmic divergence of the observed de Haas van Alphen mass; the key point is the field evolution of the topological change which has occurred at the onset of antiferromagnetism via the modification in the balance between majority and minority spin carriers and the spin dependences of the effective mass of the electrons. Quantum and topological criticalities of a Lifshitz transition have been discussed for two-dimensional correlated electron systems \cite{Yamaji2006}. Up to now, no similar study exists for the case of a three-dimensional system.

To explain the pseudo-metamagnetism in CeRu$_2$Si$_2$, a pseudo-gap model was introduced as an input parameter in the periodic Anderson model \cite{Satoh2001}. The field sweep in the pseudo-gap induces a change in sign of the exchange at the metamagnetic crossover field $H_M$ from an AF to a FM exchange. Quite recently, a phenomenological spin-fluctuation theory for an AF quantum-tricritical point (QTCP) has been developed \cite{Misawa2009}. This model is quite suitable for HFC like CeRu$_2$Si$_2$ and CeRh$_2$Si$_2$ which present both metamagnetic phenomena. Around the QTCP, both critical AF fluctuations (at $Q$) and FM fluctuations play an equivalent role in the mass enhancement. The particularity is that the singular dependence of $\gamma$ is equal to that of a conventional AF quantum critical point. In this model no power law divergence of the specific heat in three dimensions is predicted, and at the critical field $H_c$ the singular Sommerfeld coefficient $\gamma (H_c)$ is finite and given by equal footing by the AF and magnetic field-induced FM fluctuations. This is in agreement with our experimental observations reported here. The theoretical discussion takes only the singular part of the Sommerfeld coefficient $\gamma_Q$ into account. The renormalized bands lead to an additional normal contribution $\gamma_B$ through quasi-local fluctuations which is quite comparable to $\gamma_Q$.
The difference between $\gamma (p_c)$ and $\gamma (H_c)$ is reduced to a factor 1.5. Furthermore, $\gamma_B$ itself is linked to the Kondo effect (i.e. to the Kondo field $H_K$) and will decrease with magnetic field monotonously. This will push $\gamma (H_c)$ quite close to $\gamma (p_c)$. So, a reduced maximum in the field dependence of $\gamma$ at $H_c$ can be expected. From the field variation of $C/T$ at the verge to antiferromagnetism (e.g. for a concentration $x=0.1$ in Ce$_{1-x}$La$_x$Ru$_2$Si$_2$ in vicinity of the $x_c$) no clear maximum of $\gamma (H)$ is observed. However, the correction by $\gamma_B (x_c)$ may restore a maximum. Up to now a careful inelastic neutron scattering study under magnetic field for the CeRu$_2$Si$_2$ series has been performed only above $p_c$ to understand the pseudo-metamagnetic phenomena. No divergence of $\chi_Q$ was observed at $H_M$ but only a strong damping of the AF correlations \cite{Raymond2001}. According to the theory, $\chi^{-1}_Q$ will be strongly reduced far from $H_c^\star$, $(\chi^{-1}_Q \sim |H-H_c^\star |)$ while the uniform susceptibility is predicted to have only a $\sqrt{|H-H_c^\star |}$ dependence. Thus a new set of experiments with tuning the field through $H_c$ and $H_M$ just below $p_c$ on the AF side of the quantum phase transition is necessary.

For the reported Ce cases, the FM fluctuations are induced by the magnetic field. In Yb-based HFC the interplay between valence and magnetic transitions is certainly strong due to the weakness of the hybridization \cite{Flouquet2009}. FM interactions have been observed to be enhanced when both valence and magnetic fluctuations interact \cite{Watanabe2009}. The physical argument is that 
large dynamical volume fluctuations, which involve a $q \to 0$ mode, will favor the establishment of slow FM fluctuations. Thus, as observed in YbRh$_2$Si$_2$ \cite{Gegenwart2002,Knebel2006a}, one may expect a drastic change under magnetic field even for the low energy magnetic excitations \cite{Flouquet2009,Watanabe2009}. Our own view is \cite{Knebel2006a} that even in YbRh$_2$Si$_2$ no divergence of the effective mass occurs at $H_c$ (which is not a metamagnetic transition in the case of YbRh$_2$Si$_2$). We did not observe a divergence of the $A$ coefficient of the $T^2$ term of the resistivity and further, the upper temperature $T_A$ of the $T^2$ law remains finite at $H_c$. One particularity of YbRh$_2$Si$_2$ is that $\gamma_Q/\gamma_B$ is large.

The next issue appears for us to find a material where a full determination of the Fermi surface would be possible in each phase, since actually for the three  different cases considered here, the observation of large parts of the Fermi surface is still missing, notably in the PPM phase. For the other cases of highly studied HFC with initially huge values of the effective mass ($\gamma \sim 1$ Jmol$^{-1}$K$^{-2}$) close to quantum singularity as CeCu$_6$ \cite{Loehneysen2007} or YbRh$_2$Si$_2$ \cite{Gegenwart2008}, the low electronic mean free path of the first and the low value of $H_c$ for the second make very unlikely the opportunity of a direct measurement of the Fermi surface. Thus,  the stimulating challenge is clearly to observe completely the Fermi Surface of the different AF, PM, and PPM phases.

JF thanks Pr M. Imada, Y. Kuramoto, and K. Miyake for theoretical discussions. This work was performed through the support of the ANR Delice and of Euro-
magnet II via the EU contract RII3-CT-2004-506239, and the stay of JF in Osaka through the global excellence network.


\newpage

\begin{figure}
\includegraphics[width=0.75\textwidth]{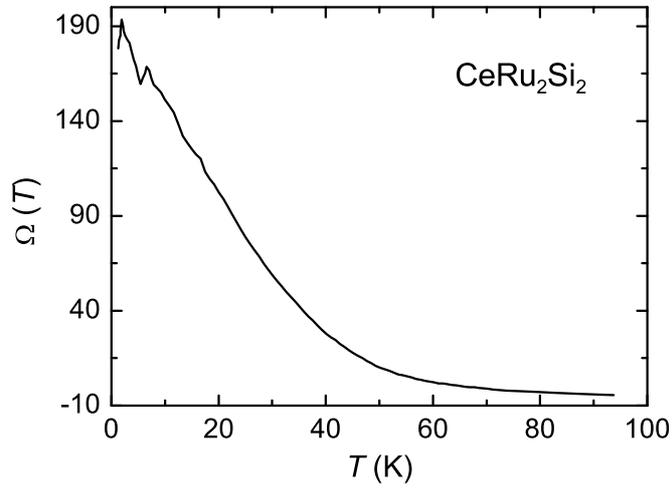}
\caption{Temperature dependence of the Gr\"uneisen parameter of CeRu$_2$Si$_2$ (after Ref. \cite{Lacerda1989}). }
\label{fig:2}
\end{figure}

\begin{figure}
\includegraphics[width=0.75\textwidth]{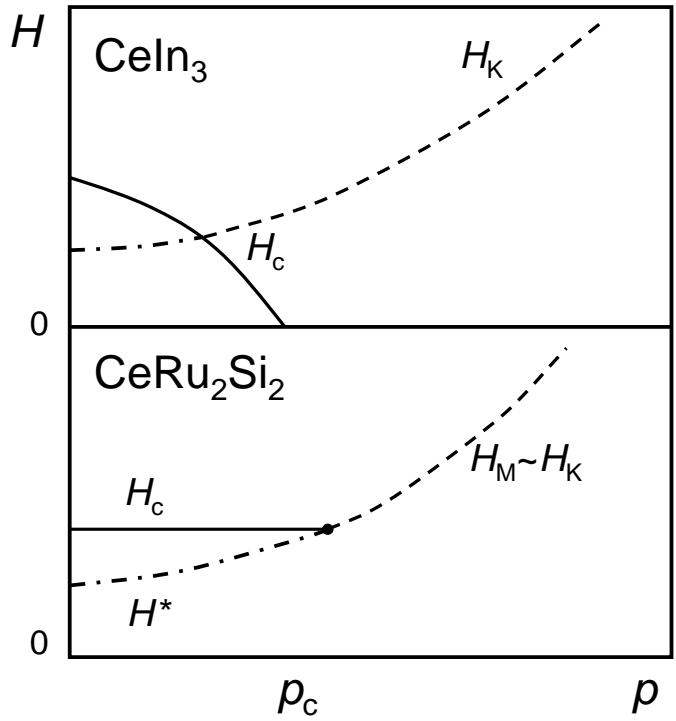}
\caption{Pressure dependence of the critical fields $H_c$ and $H_M$ for the case of CeRu$_2$Si$_2$ and CeIn$_3$ \cite{Flouquet2005}. }
\label{fig:3}
\end{figure}

\begin{figure}
\includegraphics[width=0.75\textwidth]{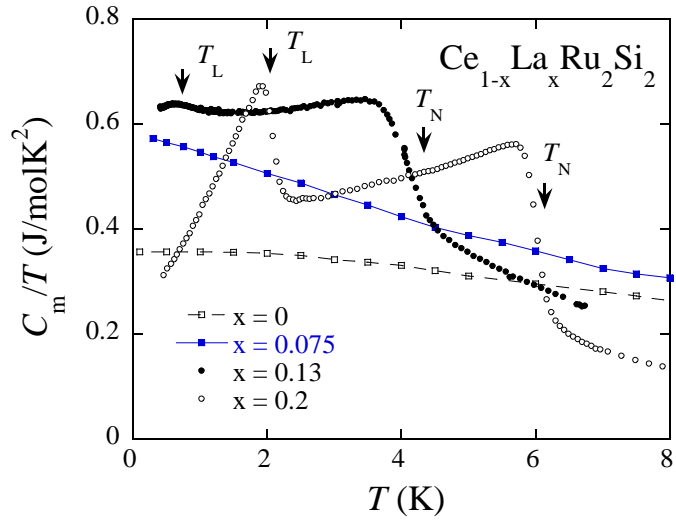}
\caption{Temperature dependence of magnetic contribution to the specific heat $C_m$ divided by temperature $T$ of Ce$_{1-x}$La$_x$Ru$_2$Si$_2$ for $x = 0$ (which corresponds to a pressure $p_c + 0.3$ GPa), $x = 0.075$ ($p_c$), $x = 0.13$ ($p_c - 0.2$ GPa), $x = 0.2$ ($p_c - 0.3$ GPa) at ambient pressure \cite{Raymond2010}. }
\label{fig:4}
\end{figure}

\begin{figure}
\includegraphics[width=0.75\textwidth]{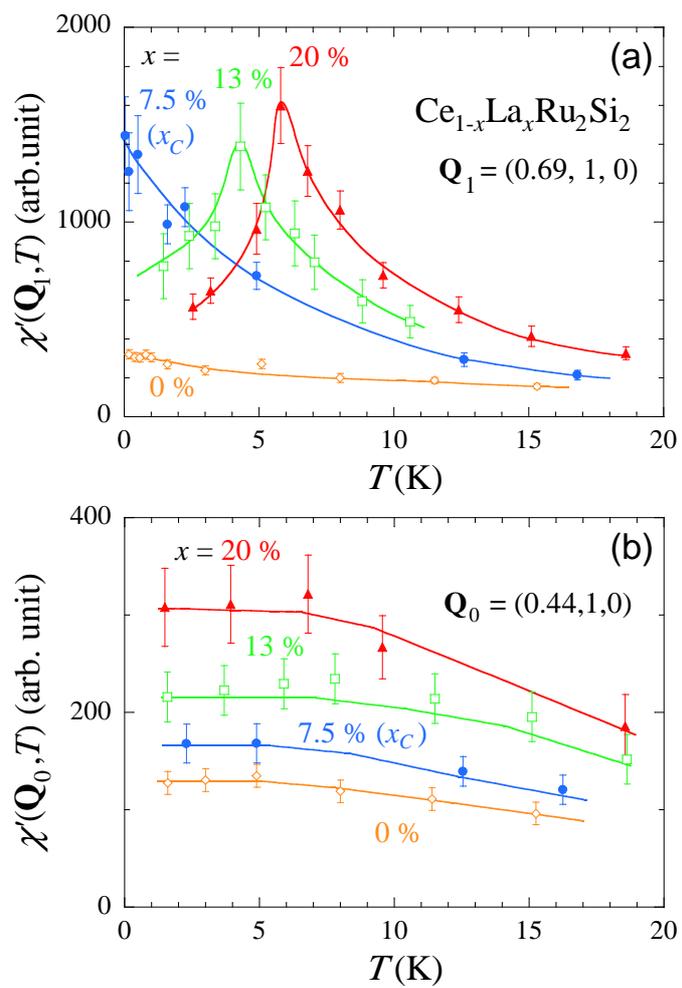}
\caption{(a) Temperature dependence of the real part of the static susceptibility $\chi ' (Q)$ at the wave vector of the AF fluctuations at $Q_1$ and (b) at a wave vector $Q = Q_0$ very far from the AF wavevector \cite{Knafo2009}.}
\label{fig:5}
\end{figure}

\begin{figure}
\includegraphics[width=0.75\textwidth]{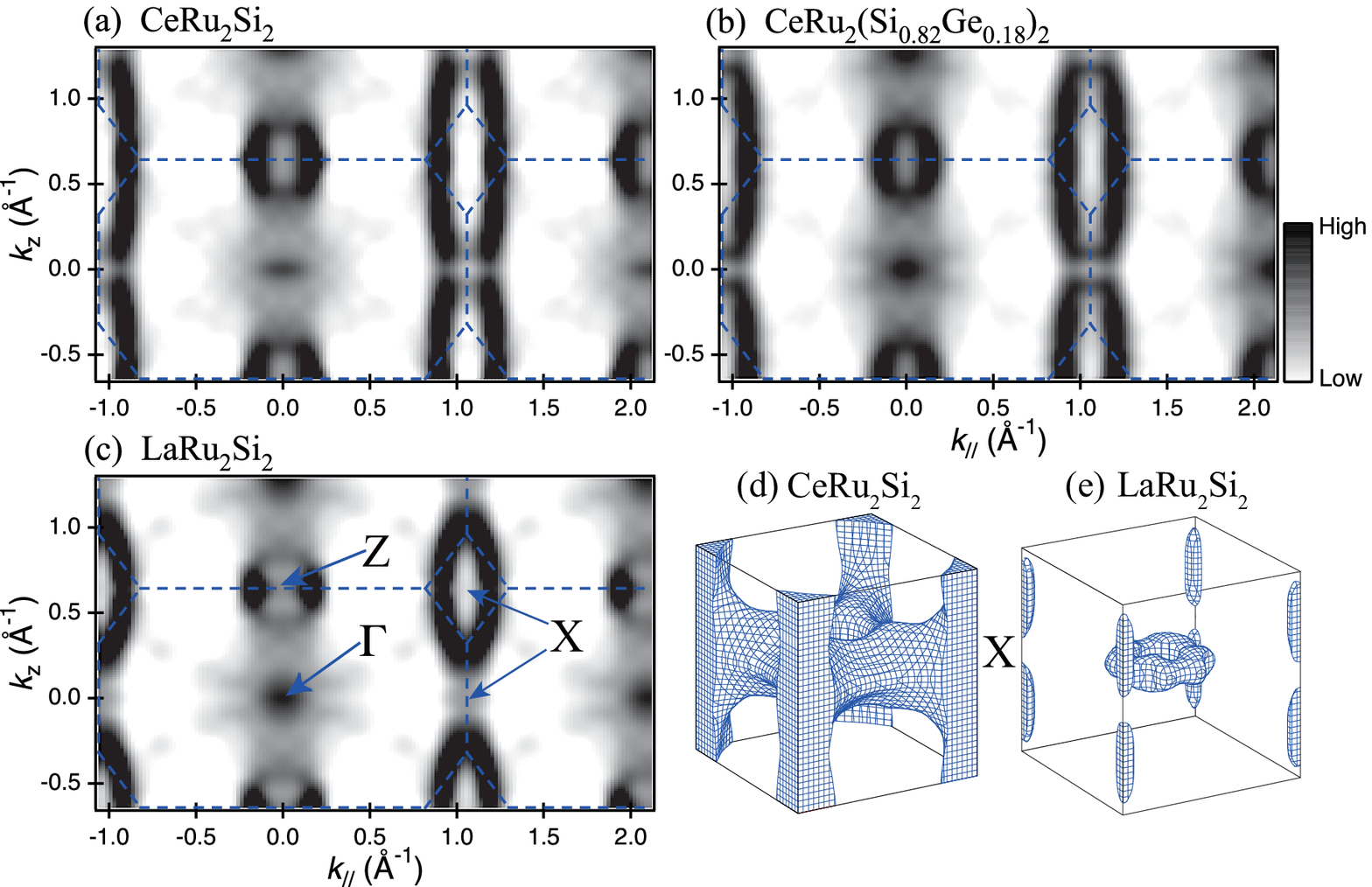}
\caption{(a-c) ARPES results on CeRu$_2$Si$_2$ and CeRu$_2$(Si$_{0.82}$Ge$_{0.18}$)$_2$ compared with LaRu$_2$Si$_2$. (d,e) Fermi surface calculated for CeRu$_2$Si$_2$ and LaRu$_2$Si$_2$. (after Ref. \cite{Okane2009})}
\label{fig:6}
\end{figure}

\begin{figure}
\includegraphics[width=0.75\textwidth]{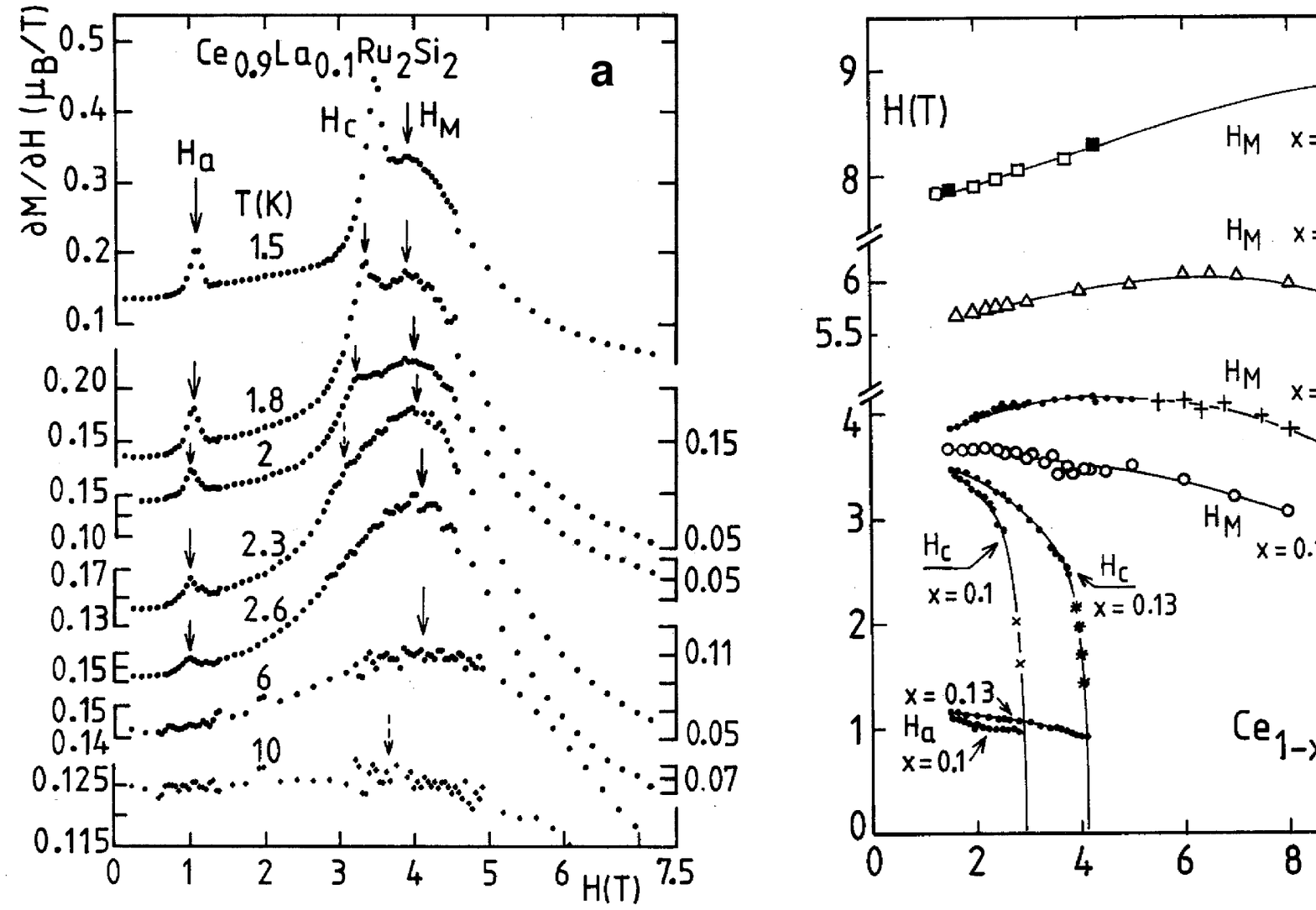}
\caption{a) Differential susceptibility of Ce$_{0.9}$La$_{0.1}$Ru$_2$Si$_2$ at different temperatures as a function of the applied field. b) Temperature variation of the critical metamagnetic field  $H_a$ and $H_c$ and of the pseudo-metamagnetic field $H_M$ \cite{Fisher1991}.}
\label{fig:7}
\end{figure}

\begin{figure}
\includegraphics[width=0.75\textwidth]{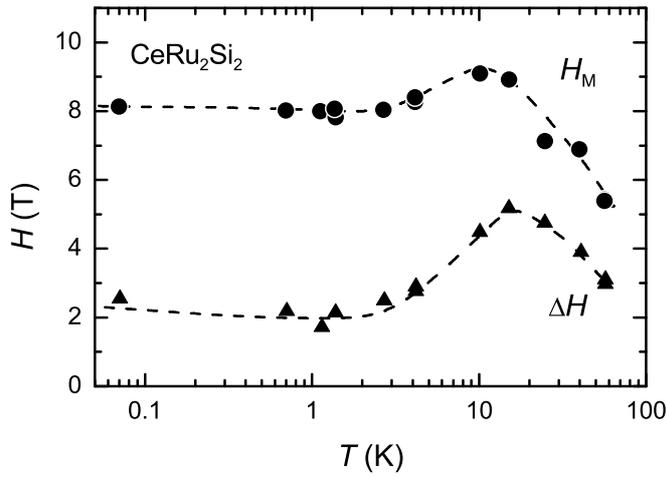}
\caption{Temperature dependence of $H_M$ for CeRu$_2$Si$_2$. $H_M$ is derived from the maximum of the positive magnetoresistance measured at constant temperature.  $\Delta H$ gives the broadening of the maximum. (see Ref. \cite{Haen1987}))}
\label{fig:8}
\end{figure}

\begin{figure}
\includegraphics[width=0.75\textwidth]{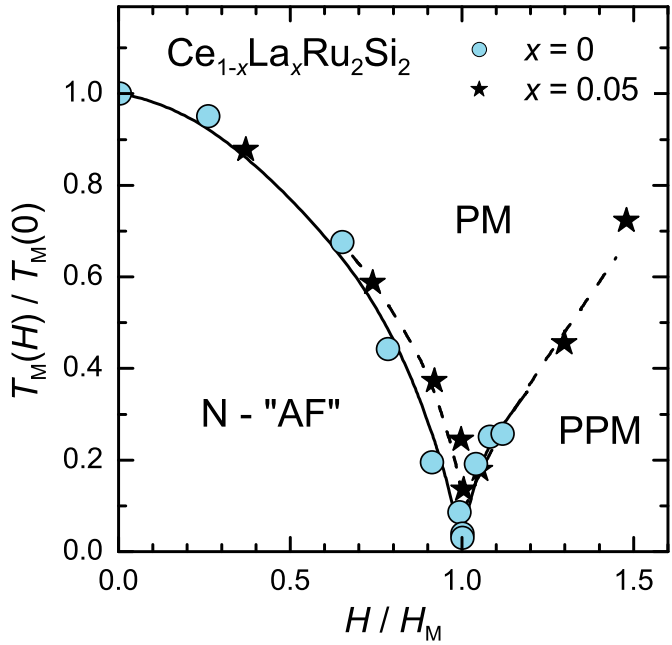}
\caption{Crossover boundaries of the different regimes of the PM phase of Ce$_{1-x}$La$_x$Ru$_2$Si$_2$ with at low field the nearly AF domain (N-''AF'') and at high field the polarized paramagnetic phase PPM. \cite{Flouquet2005,Haen1987,Paulsen1990}}
\label{fig:9}
\end{figure}

\begin{figure}
\includegraphics[width=0.75\textwidth]{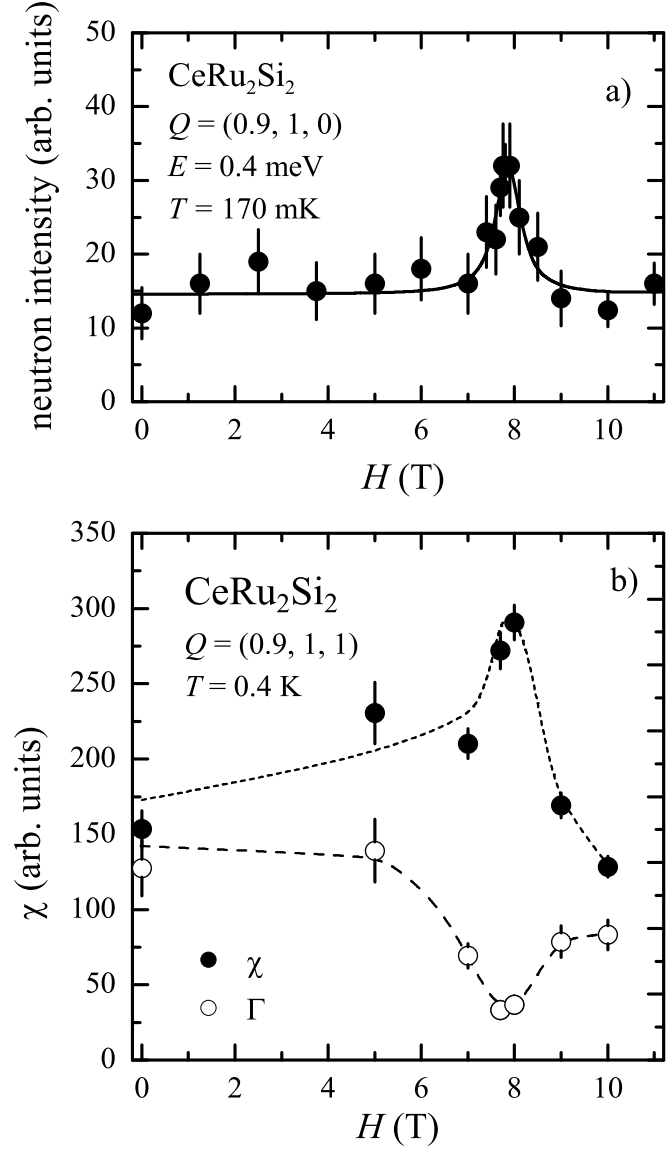}
\caption{Evidence of the softening of the FM fluctuations at $H_M$ for $Q=(0.9, 1, 0)$. a) Field dependence of the neutron intensity measured at $T=0.4$ K for an energy of $E = 0.4$ meV \cite{Flouquet2004}. b) Direct determination of the softening via the width $\Gamma$ detected in the inelastic spectrum and field variation of $\chi (0)$ which is in difference to the uniform susceptibility strongly enhanced by the huge magnetostriction at $H_M$ (taken from Ref. \cite{Sato2004}).}
\label{fig:10}
\end{figure}

\begin{figure}
\includegraphics[width=0.75\textwidth]{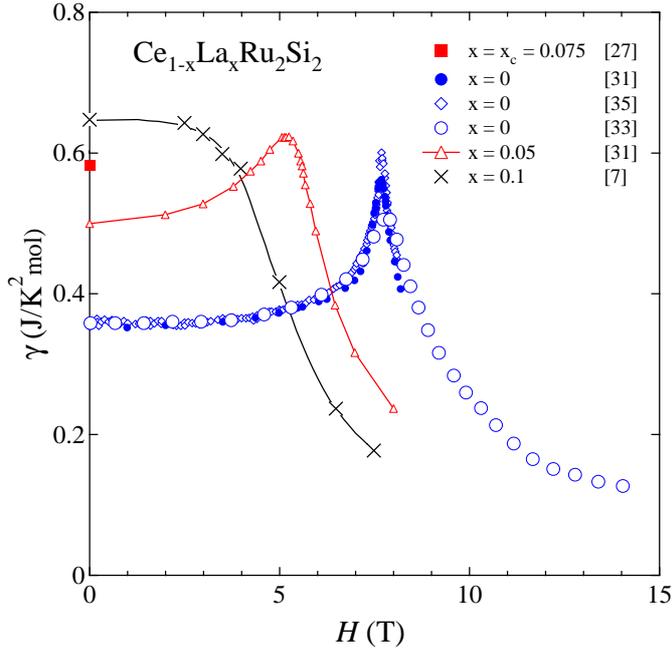}
\caption{Field variation of $\gamma=C/T$ of CeRu$_2$Si$_2$ and Ce$_{1-x}$La$_x$Ru$_2$Si$_2$. $\gamma (p_c)$ is measured at $x=x_c$ at $p=0$. For $x=0$ see Refs. \cite{Paulsen1990,vanderMeulen1991,Kim1990}, for $x=0.05$ Ref. \cite{Paulsen1990}, for $x=0.1$ Ref. \cite{Fisher1991}, and for $x=x_c=0.075$ Ref. \cite{Raymond2010}.}
\label{fig:11}
\end{figure}

\begin{figure}
\includegraphics[width=0.75\textwidth]{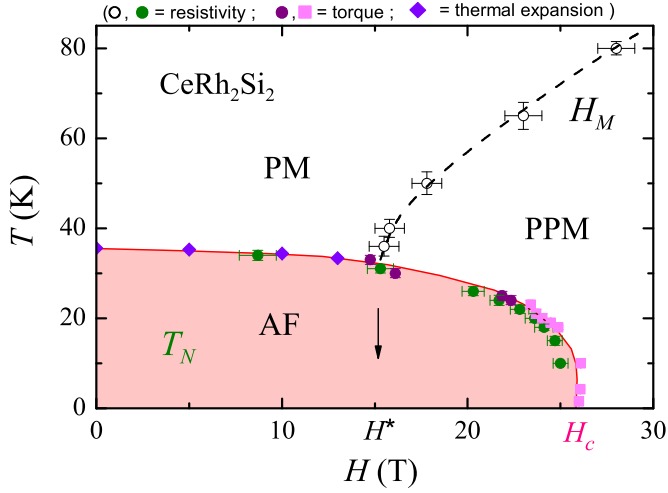}
\caption{ $(H, T)$ phase diagram of CeRh$_2$Si$_2$ at $p=0$. For clarity we have withdrawn the different phases which occur in the AF phase. The PPM boundary in absense AF order seems to end up at $H^\star \sim 15$ T for $T \to 0$ K, while the critical field to suppress the order is $H_c(0) = 27$ T. $A(H)$ is strongy enhanced above $H^\star$.  }
\label{fig:12}
\end{figure}

\begin{figure}
\includegraphics[width=0.75\textwidth]{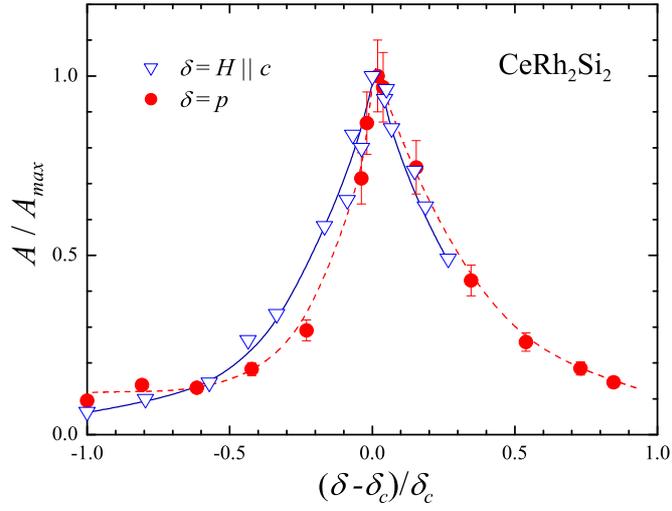}
\caption{Enhancement of the $A$ coefficient of the resistivity under pressure compared to the enhancement under magnetic field on a normalized field scale for CeRh$_2$Si$_2$ \cite{Knafo2010}.  }
\label{fig:13}
\end{figure}

\begin{figure}
\includegraphics[width=0.75\textwidth]{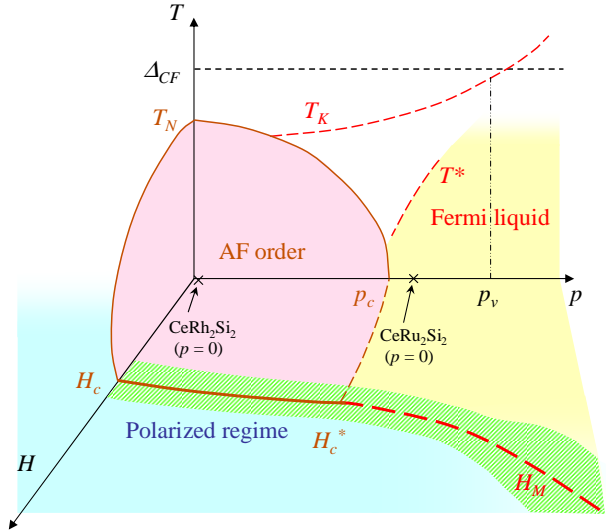}
\caption{($H, p, T$) phase diagram of CeRh$_2$Si$_2$. The dashed area correspond to the domain where the FM component will play a major role in the enhancement of $\gamma$.}
\label{fig:14}
\end{figure}

\begin{figure}
\includegraphics[width=0.75\textwidth]{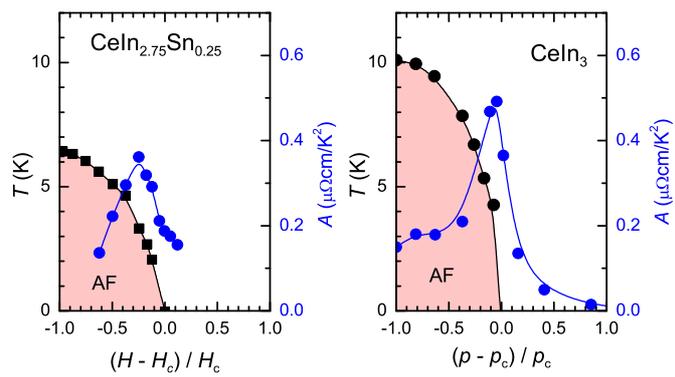}
\caption{a) $(T, H)$ phase diagram of CeIn$_{2.75}$Sn$_{0.25}$. The doping is used to lower $H_c$ from 60 T for the pure compound CeIn$_3$ to 42 T with Sn doping. This allows to determine $A(H)$ in the vicinity of $H_c$ \cite{Silhanek2006}. b) ($p, T$) phase diagram of CeIn$_3$ \cite{Knebel2001}. }
\label{fig:15}
\end{figure}

\end{document}